\documentstyle[a4,12pt,amstex]{article}

\title{Asymptotic Behavior\\ in \\
       Polarized {\bf T}$^2$-symmetric Vacuum Spacetimes}
\author{James Isenberg\thanks{Department of Mathematics and %
        Institute for Theoretical Science,
        University of Oregon, Eugene, OR 97403, USA.
        E-mail: {\tt jim@newton.uoregon.edu}}\, and  
        Satyanad Kichenassamy\thanks{%
        Max-Planck-Institut f\"ur Mathematik in den
Natur\-wis\-sen\-schaf\-ten,
        Insel\-stra\ss e 22-26, D-04103 Leipzig, Germany.
        {\em Permanent address:} Laboratoire de Math\'ematiques,
        Universit\'e de Reims, Moulin de la Housse, B. P. 1039,
        F-51687 Reims Cedex 2, France}
}
\date{}

\newcommand{\x}{\mbox{{\bf x}}}
\renewcommand{\k}{\kappa}
\newcommand{\ep}{\varepsilon}
\newcommand{\pa}{\partial}
\newcommand{\z}{\zeta}
\renewcommand{\t}{\theta}
\newcommand{\T}{{\bf T}$^2$}
\newtheorem{thm}{Theorem}

\begin{document}

\hfill To appear in: {\bf Journal of Mathematical Physics}

\vskip 3em

\begin{center}
\vbox{\Large Asymptotic Behavior

in 

       Polarized {\bf T}$^2$-symmetric Vacuum Spacetimes}

\vskip 1em

{\large James Isenberg\footnote{Department of Mathematics and
        Institute for Theoretical Science,
        University of Oregon, Eugene, OR 97403, USA.
        E-mail: {\tt jim\char'100 newton.uoregon.edu}}\, and
Satyanad Kichenassamy\footnote{%
        Max-Planck-Institut f\"ur Mathematik in den Natur\-wissenschaften,
        Inselstra\ss e 22-26, \mbox{D-04103 Leipzig,} Germany.
        E-mail: {\tt kichenas\char'100 mis.mpg.de}.
        {\em Permanent address:} Laboratoire de Math\'ematiques,
        Universit\'e de Reims, Moulin de la Housse, B. P. 1039,
        F-51687 Reims Cedex 2, France.
}}
\end{center}

\vskip 3em

\smallskip

\begin{abstract}
  We use the Fuchsian algorithm to study the behavior near the
  singularity of a class of solutions of Einstein's vacuum equations.
  These solutions admit two commuting spacelike Killing fields like
  the Gowdy spacetimes, but their twist does not vanish.  The
  spacetimes are also polarized in the sense that one of the
  `gravitational degrees of freedom' is turned off. Examining an
  analytic family of solutions with the maximum number of arbitrary
  functions, we find that they are all asymptotically velocity-term
  dominated as one approaches the singularity.
\end{abstract}

{\bf PACS numbers:} 04.20Cv, 04.20Ha, 04.20Ex

{\bf Short title:} Singular spacetimes

\vfill

\clearpage

\section{Introduction}

There is increasing evidence that cosmological solutions exhibit
rather special dynamical behavior in the neighborhood of their
singularities. The evidence is still essentially limited to families
of solutions with at least one Killing field. However, it is quite
striking that although the Hawking-Penrose singularity theorems
\cite{hawking-penrose} require nothing more than geodesic
incompleteness in generic cosmological solutions, every study to date
indicates that the solutions under investigation are either
`asymptotically velocity-term dominated' (AVD) or show `Mixmaster'
behavior,
see \cite{w-h,bkl,rendall-mix,i-m,c,gowdy,weaver-berger-isenberg}.

In a space with AVD behavior, the metric tensor $g_{ab}(\x,t)$ evolves
in such a way that an observer with fixed $\x_0$ moving toward the
singularity sees the dynamics of $g_{ab}(\x_0,t)$ asymptotically
approach that of a Kasner spacetime, with there being generally a
different Kasner limit for each different $\x_0$ (see
\cite{e-l-s,bkl,i-m,gowdy}, and references
therein). Mixmaster behavior is similar, except that this observer
sees $g_{ab}(\x_0,t)$ move through an infinite sequence of Kasner
epochs, with regular intermittent bounces from one epoch to another
(see \cite{h-b-c}).  Again, different observers generally see
different sequences (see for instance
\cite{bkl,weaver-berger-isenberg}). While neither AVD nor Mixmaster
behavior as described above is trivial, the Einstein equations, even
with the simplification of an assumed symmetry, are sufficiently
complicated that the prevalence of these special behaviors is quite
remarkable.

The earliest verifications of AVD behavior in a family of
inhomogeneous solutions, the polarized Gowdy spacetimes, took the form
of a theorem \cite{i-m,c-i-m}. The techniques developed in proving that
result have not, however, been readily extended to more general
families.  Instead, most of the recent evidence for AVD and Mixmaster
behavior in cosmological spacetimes has been based on numerical work:
Berger and Moncrief \cite{berger-moncrief} provide strong numerical
evidence for AVD behavior in general ({\bf T}$^3$) Gowdy spacetimes,
but find that the Kasner exponents should satisfy some inequalities in
generic solutions (the solutions should be `low-velocity'); they also
have evidence in polarized {\bf U}(1)-symmetric spacetimes
\cite{b-m-u1}. One should note that it is not always easy to be sure,
in numerical computations, that
the constraint equations do hold, except in the Gowdy class. Note also
that Weaver, Berger and Isenberg \cite{weaver-berger-isenberg} provide
similar evidence that locally {\bf T}$^2$-symmetric spacetimes with
certain magnetic fields have Mixmaster behavior.

This numerical evidence motivates the search for a theoretical
explanation for the prevalence of these behaviors and numerical
observations such as the distinction between high- and low-velocity
solutions, and if possible a means to predict which behavior occurs.
The recent work of Kichenassamy and Rendall \cite{gowdy} introduces a
new tool for obtaining such information. They use the Fuchsian
algorithm to prove that there is a family of general (non-polarized)
Gowdy spacetimes parametrized by the maximum number of free functions,
namely four, which all exhibit AVD `low-velocity' behavior. If the
derivative of one of these functions vanishes, `high-velocity'
behavior is allowed. This family of solutions includes all of the
previously known solutions in this class. The results also shed new
light on other features of the numerical computations.

It is very likely that one can show that these new Gowdy spacetimes are
stable under smooth perturbation of Cauchy data, by adapting the
techniques described in \cite{inversibilite,syd}. The general strategy
consists in showing, using the Nash-Moser implicit function theorem,
that the free functions which determine the solutions given by the
Fuchsian algorithm can be used to parametrize solutions much in the same
way as one uses Cauchy data on a hypersurface to label regular
solutions. In a sense, one therefore generates systematically an
`asymptotic phase space' for families of solutions, as was called for in
\cite{i-m}. 

In this work, we show that the Fuchsian algorithm is an effective tool
for proving that AVD behavior occurs in a wider class of spacetimes:
those which possess, like the Gowdy spacetimes, a {\bf T}$^2$ isometry
group with spacelike generators, but in which, unlike the Gowdy case,
the Killing vectors have a non-vanishing twist. The main new
difficulty is that this non-vanishing twist prevents the constraint
equations from decoupling from the evolution equations, resulting in a
considerably more complicated PDE system than what obtains in the case
of Gowdy spacetimes \cite{chrusciel,b-c-i-m,rendall-cmc}. This
difficulty is overcome by abandoning the separation of constraint and
evolution equations. It is found that, combining some of the
constraints with some of the `evolution' equations, one can form a
system which is sufficient to determine the metric. One then proves
directly that the remaining constraints hold everywhere if they hold
asymptotically at the singularity. This latter condition can be
expressed explicitly in terms of the data which determine the
asymptotics at the singularity, or `singularity data' for short.

The Fuchsian algorithm has been extensively studied and takes a
variety of forms (see \cite{syd,nlw,gks,gowdy}). In section II, we
briefly review the form of the algorithm we use here, and a few
relevant results we will need.  Next, we describe in section III the
{\bf T}$^2$-symmetric spacetimes, noting some of their properties and
defining the polarized sub-family. Then, in section IV, we propose an
AVD Ansatz for the metric coefficients and show that the `regular
part' of the field is indeed negligible in comparison with the leading
terms. Finally, we discuss in section V our conclusions and plans for
future work.


\section{The Fuchsian algorithm}

The Fuchsian algorithm was initially developed to understand the
behavior of solutions of differential equations in the neighborhood of
a possible singularity of unknown location. The rationale was that if
singularities are to form, it would be desirable to figure out by what
mechanism they form: which components of the solution becomes singular?
do singularities occur only in higher derivatives? is the locus of the
singularity arbitrary? how does it vary with Cauchy data given on a surface
where the solution is smooth?

Existing results prior to Fuchsian techniques gave some information on
the time of the first singularity, but did not shed light on the
mechanism of singularity formation, except for special classes of
singularities, such as shock waves in low dimensions, or caustic
formation.

The questions asked above would be answered if it were possible to
establish an expansion of the solution to relatively high order. To
achieve this, one needs to establish a formal solution, and to prove that
this formal expansion does characterize the solution. In practice,
one is not primarily interested in the convergence or
divergence of a series representation. Rather, one would like to know
whether the parameters entering in a formal series representation do
{\em determine uniquely} the solution, or whether there are infinitely
many solutions differing from each other by, say, exponentially small
corrections.

The Fuchsian method tackles this problem by seeking a reduction of the
given system of PDEs to a Fuchsian system; that is, one which has a
regular singular point with respect to one of the variables, which we
call $t$. Using a change of coordinates if necessary, one may assume
that the locus of the singularity is $t=0$. It is also possible to set
things up so that one always deals with {\em first-order} Fuchsian
systems, by the introduction of new variables. This is a familiar
procedure for the Cauchy problem: for instance, if $u$ solves the wave
equation in Minkowski space, it is easy to check that the quantities
$(u,\pa_a u)$ satisfy a first-order system.

Let us consider a PDE system which we write symbolically as
\[
F[u]=0.
\]
The exact form of the nonlinearity is not important for what follows.
Generally, $u$ can have any number of components.

Schematically, the Fuchsian algorithm has three parts:

{\bf Step 1.} Identify the {\em leading part} of the desired expansion
for $u$. This can be done in many cases by seeking a leading balance;
that is, a leading term $a(t)$ such that, upon substitution into the
equation, the most singular terms cancel each other.

{\bf Step 2.} Introduce a {\em renormalized unknown}. This means that
one writes
\begin{equation}
\label{1I}
    u=a(t)+t^s v(t),  
\end{equation}
where $v$ is the new unknown. It is generally useful to compute $a$ to
relatively high order if possible, so that any arbitrary functions in
the expansion are already included in $a$. If $a$ is a solution up to
order $n$, one may usually take $s=n+\ep$, where $\ep$ is small.

{\bf Step 3.} Obtain and solve a {\em Fuchsian system} for
$v$. Indeed, one finds under rather general circumstances that the
function $v$ solves an equation of a very particular form, namely
\begin{equation}
\label{2I}
(t\pa_t+A)v=t^\ep f(t,x^\rho,v,\nabla_\rho v),
\end{equation}
where $\rho$ stands for spatial indices in this formula. The matrix
$A$ depends at most on the spatial variables, and $f$ is, say,
bounded. By taking $t^\ep$ to be a new time variable, one may always
assume that $\ep$ is equal to one. Observe how spatial derivatives are
effectively switched off from the equation when $t$ goes to zero: {\em
  the Fuchsian algorithm provides a systematic procedure to guarantee
AVD behavior}.

There is a variety of existence results for Fuchsian systems
\cite{nlw,hs,gowdy}. For our purposes, it suffices to note the
following (see \cite{gowdy,nlw})
\begin{thm}
  There is a unique local solution which is continuous in time and
  analytic in space, and vanishes as $t$ goes to zero, provided that
  (a) $f$ is continuous in $t$ and analytic in its other arguments
  and satisfies an estimate of the form 
\[
|f(t,x^\rho,v,\nabla_\rho v)-f(t,x^\rho,w,\nabla_\rho w)|
\leq C [ |v-w|+|\nabla_\rho v-\nabla_\rho w|] 
\]
  for some constant $C$ provided $v$ and $w$ are bounded,
  and (b) the matrix
  $\sigma^A (=\exp(A\ln\sigma))$ is uniformly bounded for
  $0<\sigma<1$.
\end{thm}
Condition (b) is usually most
conveniently checked by simply computing the matrix exponential.

We emphasize that we are not allowed to prescribe arbitrarily the
initial value of $v$. The free data (usually called `singularity data')
which label the solution $u$ are already built into the choice of $a$ in
(\ref{1I}), and are subsequently incorporated into the function $f$ in
(\ref{2I}). A straightforward extension of the theorem can be made if we
assume only that $v(0)$ belongs to the null-space of $A$. By considering
the equation satisfied by $v-v(0)$, one can reduce the problem to an
equation to which the theorem applies. In such a case, $v(0)$ must be
added to the list of singularity data. 

General strategies for carrying out the algorithm can be found in
\cite{syd,gks,nlw,gowdy}, with applications to several examples. Let
us simply describe here what these steps entail for the first PDE to which
these ideas were applied successfully:
\begin{equation}
\label{3I}
\eta^{ab}\pa_{ab}u=e^u
\end{equation}
in Minkowski space, where $u$ is a scalar field (there are similar
results for power nonlinearities as well). Let $t=\psi(x)$ be the
locus of the (yet unknown) singularity, and let $x$ stand for the
spatial variables. For one space dimension, equation (\ref{3I}) has a
closed-form solution (``Liouville field theory''); however, we allow
here the number of space dimensions to be arbitrary. Let
$T=t-\psi(x)$. If we choose the leading part of $u$ so that $\exp(u)
\sim \varphi(x)T^s$ where $s$ and $\varphi$ are unknown, one readily
finds that, to eliminate the most singular term in the expansion of
(\ref{3I}), we need to choose $s=-2$ and
$\varphi=2(1-|\nabla_x\psi|^2)$ which must therefore be positive.
Hence the leading part of $u$ takes the form
\[
u\approx \ln {2\over T^2} + \ln (1-|\nabla_x\psi|^2).
\]
This completes the first step. 
 
It is useful to write out the rest of the leading part $a(t)$ of $u$
up to order two in $T$ for two reasons: (a) this reveals that the
solution contains logarithmic terms, which disappear in fact only if
the scalar curvature of the singularity manifold vanishes identically;
(b) this shows that the coefficient of $T^2$ in the expansion is
arbitrary. We therefore compute, by direct substitution
\[
u\approx \ln {2\over T^2} +u_0(x) + u_1(x)T + u_{1,1}(x)T^2\ln T +
u_2(x)T^2+\dots,
\]
where $u_0$, $u_1$ and $u_{1,1}$ are entirely determined by $\psi$;
in particular, $u_0=\ln (1-|\nabla_x\psi|^2)$. However, $u_2$ remains 
arbitrary. One then sets
\begin{equation}
  \label{4I}  
u=\ln {2\over T^2} +u_0(x) + u_1(x)T + u_{1,1}(x)T^2\ln T +
vT^2
\end{equation}
so that the arbitrary function $u_2$ appears as an `initial value' for
the renormalized unknown $v$. This completes the second step.

The singularity data in this case are $\psi$ and $u_2=v(0)$.  Once they
are known, the formal solution is completely determined.

For the third step, we now substitute expression (\ref{4I}) for $u$
into (\ref{3I}), and find that $v$ solves an equation which can be
thought of as a non-linear perturbation of the Euler-Poisson-Darboux
equation. One then checks that $v$, $Tv_T$ and $T\nabla_\rho v$
solve a Fuchsian system. This has the following consequences:

(a) There is a formal solution to all orders, in powers of $T$ and
$T\ln T$; for $T<0$, one replaces $T\ln T$ by $T\ln |T|$.  The series
are convergent if $\psi$ and $u_2$ are analytic; otherwise, they are
valid as far as the differentiability of the free functions allows. As
already mentioned, the logarithmic terms cannot be dispensed with,
except for very special solutions. The example of Gowdy spacetimes
shows that one should even allow terms such as $T^{k(x)}$ for the
application to Einstein's equations.

(b) The solution is uniquely determined by the `singularity data'
$\{\psi,u_2\}$. In fact, the correspondence between these data and the
Cauchy data on a hypersurface where the solution is regular can be
inverted, using the Nash-Moser version of the inverse function
theorem, see \cite{inversibilite}. From a practical viewpoint, this
means that (i) the smoother the Cauchy data, the smoother the
singularity surface; (ii) there is a recipe for computing how the
singularity data vary when the Cauchy data vary. The question in the
opposite direction is simpler: the series representation gives
explicitly the solution in terms of the singularity data.

All this makes the series representations generated by Fuchsian
techniques as reliable as an exact solution in the vicinity of the
singularity---a place where at present no other representation is
available.

It should be stressed that the Fuchsian method applies without
symmetry or integrability restrictions. For this reason, it enables
one to study directly the stability of solutions furnished by
generation techniques, even under fully `inhomogeneous' or
`asymmetric' perturbations, and the information it provides yields
concrete analytical insight into the properties of solutions.


\section{Polarized {\bf T}$^2$-symmetric spacetimes}

While the Gowdy {\bf T}$^3$ spacetimes \cite{g} have been extensively
studied over the years
\cite{i-m,chrusciel,c,berger-moncrief,gowdy},
and are relatively well-understood, the more general {\bf
  T}$^2$-symmetric spacetimes have only recently begun to be
considered \cite{chrusciel,b-c-i-m,rendall-cmc}. The technical
condition which distinguishes the Gowdy sub-family is the requirement
that the Killing fields $X$ and $Y$ which generate the isometry group
have vanishing twist constants $\k_x:=\ep^{abcd }X_aY_b\nabla_cX_d$
and $\k_y:=\ep^{abcd}X_aY_b\nabla_cY_d$, where $\ep^{abcd}$ is the
Levi-Civita tensor. The essential difference in practice is that, if
one chooses the constant orbit area time-foliation (``Gowdy time'')
\cite{g}, the constraint equations decouple from the evolution
equations in the Gowdy case, and can therefore more or less be ignored
in the analysis. If however either $\k_x$ or $\k_y$ is nonzero, then
no such decoupling occurs.

The general form of the metric and the field equations for the {\bf
  T}$^2$-symmetric spacetimes is presented in \cite{b-c-i-m}, along
with a proof that the Gowdy time  always exists globally for these
spacetimes. To write the metric, we assume that all metric components
depend on two coordinates, the Gowdy time $t$ and spatial coordinate
$\t\in S^1$ with $\pa/\pa x$ and $\pa/\pa y$ generating the {\bf
  T}$^2$ isometry. By choosing $X$ and $Y$ to be suitable linear
combinations of the generators, we may always assume without loss of
generality that $\k_x=0$. We then drop the subscript from $\k_y$. We
now focus our attention on the sub-class of polarized spacetimes,
which have $A\equiv 0$ in the notation of \cite{b-c-i-m}.  The metric
takes the form
  \begin{equation}
    \label{metric}
    ds^2 = e^{2(\nu-u)}(-\alpha dt^2 + d\t^2)
+\lambda e^{2u}(dx + G_1 d\t+M_1 dt)^2
+\lambda e^{-2u}t^2(dy + G_2 d\t+M_2 dt)^2
  \end{equation}
where $\lambda$ is a positive constant and the functions $u$, $\nu$,
$\alpha$, $G_1$, $M_1$, $G_2$, $M_2$ depend on $t$ and $\t$. 
The vacuum field equations take the form, writing $u_t$ for $\pa u\over
\pa t$ etc., $D=t\pa_t$, $D^2=t^2\pa^2_t + t\pa_t$, and $m=\lambda \k^2$, 
\begin{equation}  \label{7}
\begin{align}
  D^2 u - t^2\alpha u_{\t\t} & = 
            {1\over 2\alpha}D\alpha Du + {t^2\over 2}\alpha_\t u_\t
  \tag{\ref{7}a}  \\   
  D\alpha & =  -{\alpha^2\over t^2}me^{2\nu}
  \tag{\ref{7}b}  \\   
  D\nu & = (Du)^2 +t^2 \alpha u_\t^2 + {\alpha\over 4t^2} me^{2\nu}
  \tag{\ref{7}c}  \\ 
  \pa_\t \nu & = 2u_\t Du - {\alpha_\t\over 2\alpha}
  \tag{\ref{7}d}  \\   
G_{1,t} &= M_{1,\t} \tag{\ref{7}e}  \\
G_{2,t} &= M_{2,\t}+{\k \alpha^{1/2}\over t^3}e^{2\nu} \tag{\ref{7}f}
            \\
\k_t & = 0 \tag{\ref{7}g}  \\
\k_\t & = 0 \tag{\ref{7}h}  
\end{align}
\end{equation}
\stepcounter{equation}
Note that the Gowdy case is recovered if $\k=0$, $\alpha=1$, and
$G_1=G_2=M_1=M_2=0$. Since $G_{1,t} = M_{1,\t}$, $G_1d\t+M_1dt$ is 
locally an
exact differential $d\varphi$. Replacing $x$ by $x+\varphi$, we may
assume {\em locally} that $G_1=M_1=0$. Similarly, one can set $M_2=0$ 
by redefining $y$. Since these reductions are only local and may be
incompatible with global requirements, we do not consider them further,
even though they do make the geometric `degrees of freedom' more clear.

Equations (\ref{7}) constitute an initial-value problem for the
polarized spacetimes, in which the equations (\ref{7}a-d) decouple
from the rest. They form an independent system for $\{u,\alpha,
\nu\}$. Once these three functions are known, the other equations can
be solved easily.

We note that equations (\ref{7}b-d) in particular---three of the four
equations which constitute the heart of the Cauchy problem for these
spacetimes---actually derive from the constraint equations of
Einstein's theory. Unlike the Gowdy case, the wave equation (\ref{7}a)
does not decouple from the constraints, since it contains the function
$\alpha$. We therefore take (\ref{7}a-d) as our basic equations,
treating (\ref{7}a-c) as evolution equations, and (\ref{7}d) as the
only effective constraint.

The local well-posedness of the initial-value problem away from the
singularity at $t=0$ is not quite straightforward, for we must prove
that equation (\ref{7}d) propagates. This is not an immediate
consequence of standard results because we are not using any of the
standard set-ups for the initial-value problem. It nevertheless does
hold, and this can be ascertained in two ways.  

One approach is as follows (this is basically the argument used by
\cite{b-c-i-m,chrusciel}): if we choose
$\{u,u_t,\alpha,\nu,\dots\}$ at some initial time $t_0>0$ so that
they satisfy the constraint (\ref{7}d), then we can view these as an
initial data set for the Einstein equations without any symmetry and
construct a local solution in the standard way. One then uses the
results of \cite{chrusciel} to introduce coordinates in this region so
that the metric takes the form (\ref{metric}).

We can also give a direct argument, which will be useful later.
We first deal with the analytic case, which is all we need for the
results of section IV. In view of its independent interest, we show
in the appendix how to deal with the non-analytic Cauchy problem 
as well.

Away from $t=0$, the PDE system (\ref{7}a-c) is of Cauchy-Kowalewska type. 
More precisely, we can reduce it to the following first-order system for
$(z_0,z_1,z_2,\alpha,\nu) := (u,u_t,u_\t,\alpha,\nu)$:
\begin{equation*}
\begin{align*}
\pa_t z_0 & = z_1 \\
\pa_t z_1 & = \alpha \pa_\t z_2 - {z_1\over t} 
             - {m\over 2t^3} z_1 e^{2\nu} + {1\over 2} z_2 \alpha_\t \\
\pa_t z_2 & = \pa_\t z_1 \\
\pa_t \alpha & = -{\alpha^2\over t^3} m e^{2\nu} \\
\pa_t \nu & = t z_1^2 + t\alpha z_2^2 + {\alpha\over 4t^3}me^{2\nu}.
\end{align*}
\end{equation*}
In particular, ignoring the constraint
(\ref{7}d), we obtain a unique solution of the remaining equations
by prescribing the data $\{u,u_t,\alpha,\nu,\dots\}$ for $t=t_0$. 
Now let us set
\begin{equation}
\label{ndef}
N := \nu_\t - 2u_\t Du + {\alpha_\t\over 2\alpha}.
\end{equation}
Calculating
\[
  0=D\nu_\t-\pa_\t D\nu  = DN + D(2u_\t Du - {\alpha_\t\over 2\alpha})
                                -\pa_\t D\nu,
\]
we find, using (\ref{7}a-c),
\begin{equation}
\label{n-eq}
 DN - {1\over 2\alpha}ND\alpha= 0.
\end{equation}
This is a linear {\em ordinary} differential equation for $N$ (there
are no $\t$-derivatives). Hence if we choose data
$\{u,u_t,\alpha,\nu,\dots\}$ for $t=t_0$ so that $N(t_0)=0$, the
uniqueness theorem for ODEs guarantees that $N$ is identically zero
for all time.

We therefore have proved the well-posedness of the initial-value
problem. The results of \cite{b-c-i-m} ensure that the solution
remains bounded for $t>\rho$, where $\rho\geq 0$ is independent of
$\t$. It is expected that $\rho>0$ in special cases only, such as
exact Kasner spacetimes \cite{footnote}. We are interested in
asymptotics near $t=0$. Note that Fuchsian techniques may be useful for
analyzing singularities for $t$ near $\rho>0$; however, if these solutions are
non-generic in some reasonable sense, they should not contain the
full number of free parameters, and they may be non-polarized as well. It does
appear that there are consistent asymptotics of the form
$u\approx u_0$, $\nu\approx {1\over 2}\ln(t-\rho)+\nu_0(\t)$ and 
$\alpha\approx \alpha_0(\t)(t-\rho)^{-2}$.

As far as the number of free functions in the metric is concerned, one
might expect that there will only be two, since one of the
gravitational degrees of freedom has been turned off. Indeed, while
the initial data for (\ref{7}a--c) consist of four functions
$\{u,u_t,\alpha,\nu,\dots\}$, they are constrained by one relation,
{\em viz.}\ (\ref{7}d), and, if we set aside the choice of the initial
value for the lapse function $\alpha$, we obtain two arbitrary
functions in the solution.

Similarly, we will obtain a family of {\em singular} solutions of
(\ref{7}a--c) depending on four arbitrary functions occurring in its
singular expansion, and will show that if these `singularity data' are
constrained by one relation, the constraint (\ref{7}d) holds for all
time as well.


\section{Application of the Fuchsian algorithm}

We are interested in generating solutions to (\ref{7}) which have
controlled asymptotics near $t=0$ and which are parametrized by as
many arbitrary singularity data as possible. We achieve this by
following the program outlined in section II.

\smallskip

{\bf Step 1. Leading-order asymptotics.}

Since we expect Kasner-like behavior at the singularity, and since $u$
and $\nu$ appear in the metric exponentially, we choose logarithmic
leading terms for $u$ and $\nu$:
\begin{equation}\label{9I}
\begin{align}
  u      &  \approx  k(\t)\ln t + u_0(\t) + \dots; 
\tag{\ref{9I}a} \\
  \nu    &  \approx  (1+\sigma(\t))\ln t + \nu_0(\t) + \dots; 
\tag{\ref{9I}b} \\
  \alpha &  \approx  \alpha_0(\t)+\dots.
\tag{\ref{9I}c}
\end{align} 
\end{equation}
\stepcounter{equation}
For equation (\ref{7}b) to hold at leading order, it is sufficient
that $\sigma>0$. For (\ref{7}c) to hold at leading order, one needs
$D\nu$ and $(Du)^2$ to balance each other, which requires that
\begin{equation}
\label{10I}
k^2=1+\sigma
\end{equation}
which we assume from now on. The function $\alpha_0$ should be taken
to be positive, to ensure the metric has the correct signature.

Note that there are four free functions, namely
$(k,u_0,\alpha_0,\nu_0)$, in these leading term expansions, just as
there were four Cauchy data in the discussion of section III. These four
free functions are the singularity data for this system. They are
$2\pi$-periodic; furthermore, $\alpha_0$ and $\sigma=k^2-1$ are
positive.

These asymptotics may be compared with those of the solutions obtained in
the Gowdy case in \cite{gowdy}. If $k_G$ denotes the parameter called $k$ in
\cite{gowdy}, the correspondence is: $\pm k_G = 2k-1$. This means that 
the solutions we obtain here, with $k^2>1$, 
are similar to the ``high-velocity'' Gowdy solutions, for which 
$k_G>1$. The asymptotics (\ref{9I}a-c) are not compatible with equations 
(\ref{7}) if $0<k<1$, unless $m=0$, which is the Gowdy case. Indeed, 
(\ref{7}b) implies that $\alpha$ is of the order $t^{2\sigma}$, which is
singular if $\sigma=k^2-1$ is negative. This makes the term
$D\alpha Du/(2\alpha)$ in (\ref{7}a) more singular than all the other terms
in this equation, so that (\ref{7}a) cannot hold. There are two ways to 
circumvent this: (1) take $k=0$, so that $Du$ vanishes to leading order,
giving a consistent balance, at the expense of losing the freedom to 
vary $k$; (2) add terms to the field equations which would compensate the
most singular term in (\ref{7}a)---which is possible by going over to the
non-polarized field equations. These possibilities will be addressed
when we deal with non-polarized spacetimes, in a forthcoming paper.

\smallskip

{\bf Step 2. Renormalized unknown.}

We now introduce new unknowns which will provide an exact form for the
remainders indicated with `\dots' in (\ref{9I}a-c). Because of the
$e^{2\nu}$ term, we see that it is not possible to assume that the
remainder terms are of order $t$. We do expect them to be of order
$t^\ep$ if $\ep$ is small compared to the minimum of $\sigma$. We
therefore define the renormalized unknowns $(v,\mu,\beta)$ by
\begin{equation}
  \label{ren-unkn}
\begin{align} 
  u(\t,t)       & =  k(\t)\ln t + u_0(\t) + t^\ep v(\t,t);
\tag{\ref{ren-unkn}a} \\
  \nu(\t,t)     & =  k^2(\t)\ln t + \nu_0(\t) + t^\ep\mu(\t,t);\label{ru-3}
\tag{\ref{ren-unkn}b} \\
  \alpha(\t,t)  & =  \alpha_0 + t^\ep \beta(\t,t).     \label{ru-2}    
\tag{\ref{ren-unkn}c} 
\end{align} 
\end{equation}
\stepcounter{equation}

\smallskip

{\bf Step 3. Fuchsian system.}

We shall now show that the renormalized field variables
solve a Fuchsian problem. Consequently, once the functions
$(k,u_0,\alpha_0,\nu_0)$ have been specified, and $\ep$ has been
chosen small enough, the unknowns $v$, $\mu$ and $\beta$ are uniquely
determined via theorem 1.

To achieve this, let us first, since we are looking for a first-order
system, introduce first-order derivatives of $v$ as new unknowns. This
suggests letting
\[
\vec{v} = (v_1,v_2,v_3,v_4,v_5):=(v,Dv,t^\ep v_\t,\beta,\mu).
\]
Let us also introduce the abbreviation $E=m\exp(2\nu_0+2t^\ep\mu)$. It
is helpful to remove the $t$-derivatives of $\alpha$ in the right-hand
side of (\ref{7}a) by using:
\begin{equation}
\label{12I}
{D\alpha\over \alpha} = -\alpha t^{2\sigma(\t)}E,
\end{equation}
which follows from (\ref{7}b) and (\ref{ren-unkn}c). We then
find the following evolution equations for $\vec{v}$:
\begin{equation}
  \label{sf}
\begin{align}
  Dv_1  = \mbox{}& v_2;  \tag{\ref{sf}a}\\
  Dv_2 + 2\ep v_2 + \ep^2 v_1  = \mbox{}& t^{2-\ep}(\alpha_0+t^\ep\beta)
(k_{\t\t}\ln t + u_{0,\t\t}+v_{3,\t}) \notag\\
       &   \mbox{}\quad -{1\over 2}E\alpha
                       t^{2\sigma-\ep}(k+t^\ep(v_2+\ep v_1))\notag\\
       &    \mbox{}\qquad+{1\over 2}t^{2-\ep}
    (\alpha_0+t^\ep\beta)(k_\t\ln t +u_{0,\t}+ v_3);
\tag{\ref{sf}b}\\
  Dv_3  = \mbox{}& t^\ep\pa_\t(\ep v_1+v_2);            
\tag{\ref{sf}c}  \\
  (D+\ep)v_4  = \mbox{}&-t^{2\sigma-\ep}(\alpha_0+t^\ep\beta)^2E; 
\tag{\ref{sf}d}\\
  (D+\ep)v_5  = \mbox{}& 2k(v_2+\ep v_1) +t^\ep(v_2+\ep v_1)^2
          +{1\over 4}Et^{2\sigma-\ep}(\alpha_0+t^\ep\beta)\notag\\
       &    \mbox{}\quad
+\alpha t^{2-\ep}(k_\t\ln t +u_{0,\t}+v_3)^2.
\tag{\ref{sf}e}
\end{align}
\end{equation}
\stepcounter{equation}
This system has the general form
\[
(D+A)\vec{v}=t^\ep \vec{f}(t,x,\vec{v},\pa_\t\vec{v}),
\]
where 
\[
A=\left(
\begin{array}{rrrrr}
0      & -1     &  0  &  0    &  0 \\
\ep^2  &  2\ep  &  0  &  0    &  0 \\
0      &  0     &  0  &  0    &  0 \\
0      &  0     &  0  &  \ep  &  0 \\
-2k\ep & -2k    &  0  &  0    &  \ep 
\end{array}
\right),
\]
and $\vec{f}$ is a five-component object containing all the terms in
the system that are not already included in the right-hand side.

By taking $\ep$ small (less than the smaller of 1 and any possible
value of $\sigma$), we can ensure that $\vec{f}$ is continuous in
$t$ and analytic in all the remaining variables. 
Since the eigenvalues of $A$ are $\ep$ and $0$, of multiplicities four
and one respectively, we conclude that the boundedness condition of
Theorem 1 holds. Explicitly, we have $P^{-1}AP = A_0$, hence 
$\sigma^A=P\sigma^{A_0}P^{-1}$, where
\[
A_0=\left(
\begin{array}{rrrrr}
0 & 0   & 0   & 0   & 0  \\
0 & \ep & 0   & 0   & 1  \\
0 & 0   & \ep & 0   & 0  \\
0 & 0   & 0   & \ep & 2k \\
0 & 0   & 0   & 0   & \ep
\end{array}
\right),
\mbox{ and }
P=\left(
\begin{array}{rrrrr}
0 & 1    & 0 & 0 & 0  \\
0 & -\ep & 0 & 0 & -1 \\
1 & 0    & 0 & 0 & 0  \\
0 & 0    & 1 & 0 & 0  \\
0 & 0    & 0 & 1 & 0  
\end{array}
\right),
\]
so that
\[
\sigma^{A_0}=\left(
\begin{array}{rrrrr}
0 & 0          & 0   & 0   & 0  \\
0 & \sigma^\ep & 0   & 0   & \sigma^\ep \ln\sigma  \\
0 & 0          & \sigma^\ep       & 0 & 0 \\
0 & 0          & 0   & \sigma^\ep & 2k\sigma^\ep \ln\sigma \\
0 & 0          & 0   & 0          & \sigma^\ep
\end{array}
\right).
\]
We conclude from Theorem 1 that there is a unique
solution of the Fuchsian system (\ref{sf}) which vanishes as $t$ tends
to zero, and which is analytic   in $\t$ and continuous in time. We
note in particular that if we construct $u$, $\nu$ and $\alpha$ from
(\ref{ren-unkn}a-c) with $v=v_1$, $\mu=v_5$, and $\beta=v_4$, then
$(u,\nu,\alpha)$ is a solution of equations (\ref{7}a-c). To verify
this, we note that equations (\ref{sf}a-c) imply that
\[
D(v_3-t^\ep v_{1,\t})=0,
\]
so that any solution which tends to zero with $t$ has also the
property that $v_2=tv_{1,t}$ and $v_3=t^\ep v_{1,\t}$.

We now wish to show that, by imposing a constraint on the singularity
data $(k,u_0,\alpha_0,\nu_0)$, we can guarantee that the solution
$(u,\nu,\alpha)$ of (\ref{7}a-c) obtained by solving the Fuchsian
system (\ref{sf}) will satisfy the constraint (\ref{7}d) as well, in
order to obtain genuine solutions of Einstein's vacuum equations. We
achieve this using (\ref{n-eq}), which in turn has been derived using
only (\ref{7}a--c).

First of all, since $\vec{f}$ is bounded, we know that $(D+A)\vec{v}$
is actually $O(t^\ep)$, which implies in particular that $\alpha$ and
$D\alpha$ are of order one and $t^\ep$ respectively. In particular,
$D\alpha/\alpha=t\alpha_t/\alpha=O(t^\ep)$. 
This means, using (\ref{n-eq}), that
\[
{\pa_t N\over N} = {\alpha_t\over 2\alpha}= O(t^{\ep-1})
\]
which is integrable up to $t=0$. (One could also have estimated
$D\alpha/\alpha$ directly from (\ref{12I}).) Letting $z(t,\t)$ be the
integral of this function from 0 to $t$, we find that 
\[
N(t,\t)\propto \exp z(t,\t).
\]
Thus, if we can choose the data so that $N\to 0$ as $t\to 0$ for
fixed $\t$, we will know that $N$ is in fact identically zero, and
therefore that the constraint is satisfied. Now
\begin{eqnarray*}
N & = & \nu_\t - 2u_\t Du + {\alpha_\t\over 2\alpha} \\
  & = & \nu_{0,\t} -2ku_{0,\t} + {\alpha_{0\t}\over 2\alpha_0} + o(1),
\end{eqnarray*}
where $o(1)$ is some expression which tends to zero with $t$. 
We conclude that the constraint is satisfied if and only if the
singularity data satisfy:
\begin{equation}\label{cond}
  \nu_{0,\t} -2ku_{0,\t} + {\alpha_{0\t}\over 2\alpha_0} = 0. 
\end{equation}

Note also that all the considerations in this paper are in fact local in
$\t$, and therefore allow in principle for other spatial
topologies.

To summarize, we have proved the following result:
\begin{thm}
  For any choice of the singularity data $k(\t)$, $u_0(\t)$,
  $\nu_0(\t)$ and $\alpha_0(\t)$, subject to condition (\ref{cond}),
  the \T-symmetric vacuum Einstein equations have a solution of the
  form (\ref{ren-unkn}) where $\beta$, $v$ and $\nu$ are bounded near
  $t=0$. It is unique once the twist constant $\kappa$ has been fixed,
  except for the freedom in the functions $G_1$,
  $G_2$, $M_1$ and $M_2$. Each of these solutions generates spacetimes
  with AVD asymptotics.
\end{thm}

\section{Concluding remarks}

We have therefore obtained a family of singular \T-symmetric
spacetimes with precise asymptotics at the singularity, which is of
AVD type, and which depends on the maximum number of singularity
data, that is, as many singularity data as there are Cauchy data for
solutions away from the singularity. 
Fuchsian techniques therefore apply even if the constraints do not decouple
from the `evolution' equations as in the Gowdy case.

We may also note the following.

First, it is likely that, as in the case of scalar fields, these
singular solutions are stable in a Sobolev topology, by application of
the Nash-Moser theorem, in which case these solutions form an open set
in the space of all solutions. This means that this type of AVD
behavior is {\em stable} in this class, and is therefore not a special
feature of some closed-form solution.

Second, the polarized U(1)-symmetric solutions are believed to be AVD
as well \cite{b-m-u1}, and work is underway to address this class by Fuchsian
methods.

Third, it appears that the general (nonpolarized) \T-symmetric
spacetimes may show Mixmaster behavior \cite{i-w}. Numerical
and analytical work to explore this possibility is being carried out.

\section*{Appendix}

In this appendix, we consider the non-analytic version of the
initial-value problem for \T-symmetric spacetimes. The strategy is as follows:
We first promote $\alpha_\t$ to a new field variable $\zeta :=\alpha_\t$,
and produce an evolution equation for $\zeta$ by differentiating 
(\ref{7}b) with respect to $\t$. We then use (\ref{7}b) to eliminate
$D\alpha$ from (\ref{7}a), and equation (\ref{7}d) to express $\pa_\t\nu$
in terms of the other field variables. This gives us 
a symmetric-hyperbolic system (\ref{A}) for $(z_0, z_1,z_2,\alpha,\zeta,\nu)$. 
Standard theorems then ensure that (\ref{A}) admits a unique solution,
defined in a small time interval, 
for non-analytic, but sufficiently smooth, initial data. We then show that
the constraints $\z=\alpha_\t$ and $N=0$ do propagate, by a variant of the
argument used for the propagation of the constraint $N=0$. This will establish
that we do obtain solutions to (\ref{7}a-d) non-analytic initial data.

We proceed with the details of this argument. The symmetric-hyperbolic
system is:
\begin{equation}
\label{A}
\begin{align}
\pa_t z_0 & = z_1 
                      \tag{\ref{A}a} \\
\pa_t z_1 & = \alpha \pa_\t z_2 - {z_1\over t}
             - {m\over 2t^3} z_1 e^{2\nu} + {1\over 2} z_2 \z 
                      \tag{\ref{A}b} \\
\alpha\pa_t z_2 & = \alpha\pa_\t z_1 
                      \tag{\ref{A}c} \\
\pa_t \alpha & = -{\alpha^2\over t^3} m e^{2\nu} 
                      \tag{\ref{A}d} \\
\pa_t \nu & = t z_1^2 + t\alpha z_2^2 + {\alpha\over 4t^3}me^{2\nu} 
                      \tag{\ref{A}e} \\
\pa_t \z   & = -{2m\alpha\over t^3}e^{2\nu}
     [\z+\alpha(2tz_1z_2-{\z\over 2\alpha})].   
                      \tag{\ref{A}f}           
\end{align}
\end{equation}
\stepcounter{equation}
One verifies by inspection that this system is symmetric-hyperbolic,
so that if we prescribe sufficiently smooth initial
data $\{u,u_t,\alpha,\z,\nu\}$ for $t=t_0$, we obtain
a unique solution. The first and third equations
ensure respectively that $z_1=\pa_t z_0$ and $\pa_t(z_2-\pa_\t z_0)=0$;
we may thus set $z_0=u$, $z_1=u_t$ and $z_2=u_\t$. Equations
(\ref{7}a-c) therefore hold, with $\alpha_\t$ replaced by $\z$ in 
(\ref{7}a).

Now, let us set
\begin{equation}
\label{A2}
R:=\z-\alpha_\t \mbox{ and } N' := \nu_\t - 2u_\t Du + {\z\over 2\alpha}.
\end{equation}
We proceed to derive a first-order system of ODEs for $R$ and $N'$.
For the rest of this section, we write $N$ for $N'$, for convenience.

First of all, using equations (\ref{A}d) and (\ref{A}f),
\begin{equation}\label{A3}
\begin{align}
DR &= D(\z-\alpha_\t) \notag\\
   & = -{2m\alpha\over t^2}e^{2\nu}[\z+\alpha(2u_\t Du-{\z\over 2\alpha})]
          -\pa_\t(-{\alpha^2\over t^2}me^{2\nu}) \notag\\
   & = -{2m\alpha\over t^2}e^{2\nu}[R-\alpha N] \notag\\
   & = 2{D\alpha\over \alpha}[R-\alpha N]. \tag{\ref{A3}}
\end{align}
\end{equation}
\stepcounter{equation}
Using the expression for $N$ from (\ref{A2}), taking the relation
$\z=\alpha_\t + R$ into account, we have
\[
DN = (D\nu)_\t-2DuDu_\t-2u_\t D^2u+D({\alpha_\t+R\over 2\alpha}),
\]
or
\[
DN-D({R\over 2\alpha}) =
(D\nu)_\t-2DuDu_\t-2u_\t D^2u + D({\alpha_\t\over 2\alpha}).
\]
Then, from (\ref{A}a,b,d) and the definition of $R$, we find
\begin{equation*}
\begin{align*}
DN-D({R\over 2\alpha}) 
& = \pa_\t(-{D\alpha\over 4\alpha})-{D\alpha\over \alpha}u_\t Du
      -t^2u_\t^2R + D({\alpha_\t\over 2\alpha}) \\
& = -t^2u_\t^2R + \pa_\t({D\alpha\over 4\alpha})
                -{D\alpha\over 2\alpha}(2u_\t Du).
\end{align*}
\end{equation*}
Since, from (\ref{7}b), one has
\[
({D\alpha\over 4\alpha})_\t = 
{D\alpha\over 2\alpha}(\nu_\t+{\alpha_\t\over 2\alpha}),
\]
it follows that
\begin{equation}\label{A4}
DN-D({R\over 2\alpha})+t^2u_\t^2R
      = {D\alpha\over 2\alpha}(N-{R\over 2\alpha}).
\end{equation}
Thus, combining (\ref{A3}) and (\ref{A4}), we have
\begin{equation}\label{A5}
\begin{align}
DN &= N{D\alpha\over 2\alpha} + R(D({1\over 2\alpha})-t^2u_\t^2)
       - {RD\alpha\over 4\alpha^2} + {D\alpha\over \alpha^2}[R-\alpha N]\notag\\
   &= R[{D\alpha\over \alpha^2}(-{1\over 2}-{1\over 4}+1)-t^2u_\t^2] 
       - N {D\alpha\over 2\alpha} \notag\\
   &= R[{D\alpha\over 4\alpha^2}-t^2u_\t^2]
       - N {D\alpha\over 2\alpha}. \tag{\ref{A5}}
\end{align}
\end{equation}
\stepcounter{equation}
Equations (\ref{A3}) and (\ref{A5}) constitute 
a linear, homogeneous system of ODEs for
$R$ and $N$. Therefore, if the initial data are such that these quantities 
are zero for $t=t_0$, they remain so for all time, QED.

\section*{Acknowledgments}

We thank the Max-Planck-Institut f\"ur Gravitationsphysik in Potsdam,
where this work was initiated,
for its hospitality, and we thank A. D. Rendall for helpful
discussions. Partial support for this research has come from NSF grant
PHY-9308117 at Oregon.

\end{document}